\documentclass[epj]{svjour}
\usepackage{graphics}
\begin{document}
\title{Study of the production and decay properties of neutron-deficient nobelium isotopes}

\author{M.S. Tezekbayeva\inst{1,2,a}, A.V. Yeremin\inst{1,3}, A.I. Svirikhin\inst{1,3}, A. Lopez-Martens\inst{4}, M.L. Chelnokov\inst{1}, V.I. Chepigin\inst{1}, \ A.V. Isaev\inst{1}, I.N. Izosimov\inst{1}, A.V. Karpov\inst{1,3}, A.A. Kuznetsova\inst{1}, O.N. Malyshev\inst{1,3}, R.S. Mukhin\inst{1}, A.G. Popeko\inst{1,3}, Yu.A. Popov\inst{1,3}, V. A. Rachkov\inst{1,3}, B.S. Sailaubekov\inst{1,2,5}, E.A. Sokol\inst{1}, K. Hauschild\inst{4}, H. Jacob\inst{4}, R. Chakma\inst{6}, O. Dorvaux\inst{7}, M. Forge\inst{7}, B. Gall\inst{7}, K. Kessaci\inst{7}, B. Andel\inst{8} , S. Antalic\inst{8}, A. Bronis\inst{8}, \and P. Mosat\inst{8}
%
}                     
\offprints{$^{a}$tezekbaeva@jinr.ru}          
\institute{Flerov Laboratory of Nuclear Reactions, JINR, Dubna, Russia
 \and The Institute of Nuclear Physics 050032 Almaty, The Republic of Kazakhstan
 \and Dubna State University, Dubna, Russia
 \and IJCLAb, IN2P3-CNRS, Universit\'{e} Paris Saclay, F-91405 Orsay, France
 \and L.N. Gumilyov Eurasian National University, Nur-Sultan, Kazakhstan
 \and GANIL, 14076 Caen Cedex 5, France
 \and Universit\'{e} de Strasbourg, CNRS, IPHC UMR 7178, F-67000 Strasbourg
 \and Comenius University in Bratislava, 84248 Bratislava, Slovakia}
\date{Received: date / Revised version: date}
%
\abstract{
The new neutron-deficient isotope $^{249}$No  was synthesized for the first time in the fusion-evaporation reaction $^{204}$Pb($^{48}$Ca,3n)$^{249}$No. After separation, using the kinematic separator SHELS, the new isotope was identified with the GABRIELA detection system through genetic correlations with the known daughter and granddaughter nuclei $^{245}$Fm and $^{241}$Cf. The alpha-decay activity of $^{249}$No has an energy of 9129(22)$~$keV and half-life 38.3(2.8) ms. An upper limit of 0.2\% was measured for the fission branch of $^{249}$No. Based on the present data and recent information on the decay properties of $^{253}$Rf and aided by Geant4 simulations, the ground state of $^{249}$No is assigned the 5/2$^+$[622] neutron configuration and a partial decay scheme from $^{253}$Rf to $^{245}$Fm could be established. The production cross-section was found to be $\sigma$(3n)=0.47(4) nb at a mid-target beam energy of 225.4 MeV, which corresponds to the maximum of the calculated excitation function. Correlations of the $^{249}$No alpha activity with subsequent alpha decays of energy 7728(20) keV and half-life $1.2_{-0.4}^{+1.0}$ min provided a firm measurement of the electron-capture or $\beta^{+}$ branch of $^{245}$Fm to $^{245}$Es.  The excitation function for the 1n, 2n and 3n evaporation channels was measured. In the case of the 2n-evaporation channel $^{250}$No, a strong variation of the ground state and isomeric state populations as a function of bombarding energy could be evidenced. 
} 
\authorrunning {M.S. Tezekbayeva et al.}
\titlerunning {Study of the production and decay properties of neutron-deficient nobelium isotopes}
\maketitle
\section{Introduction}
\label{intro}
\begin{figure}[!ht]
\begin{center}
\resizebox{0.45\textwidth}{!}{%
  \includegraphics{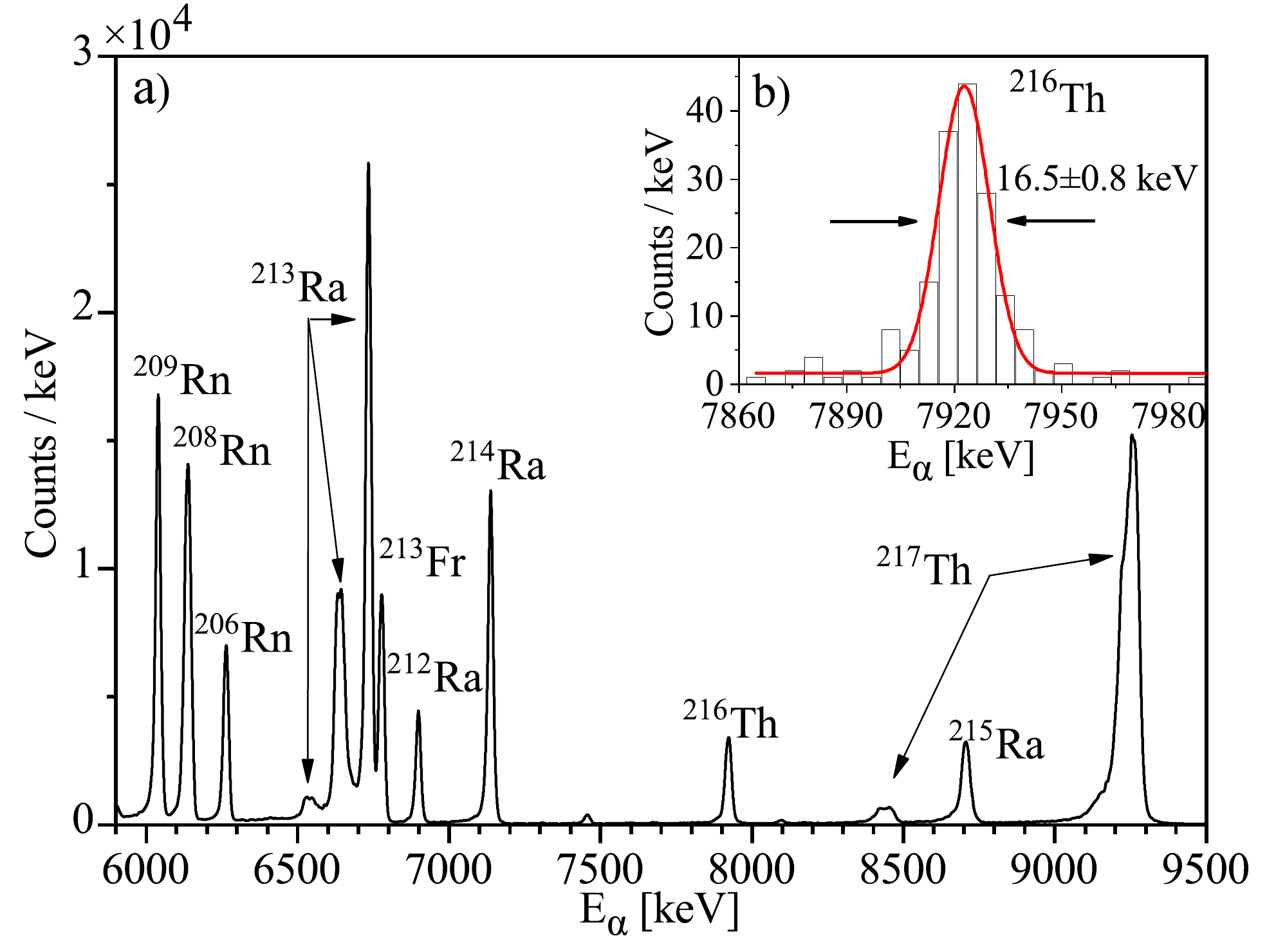}
}
\caption{a) $\alpha$ particle energy spectrum emitted by the nuclei produced in the reaction $^{48}$Ca+$^{174}Yb\rightarrow^{222}$Th$^{*}$ and detected in the implantation DSSD. b) The inset shows the fit of the $^{216}$Th alpha peak at 7922.8(3) keV from which a resolution FWHM=16.5(8) keV was extracted}
\label{fig.calib}       
\end{center}
\end{figure}
\begin{figure*}[!ht]
\begin{center}
\resizebox{0.85\textwidth}{!}{%
  \includegraphics{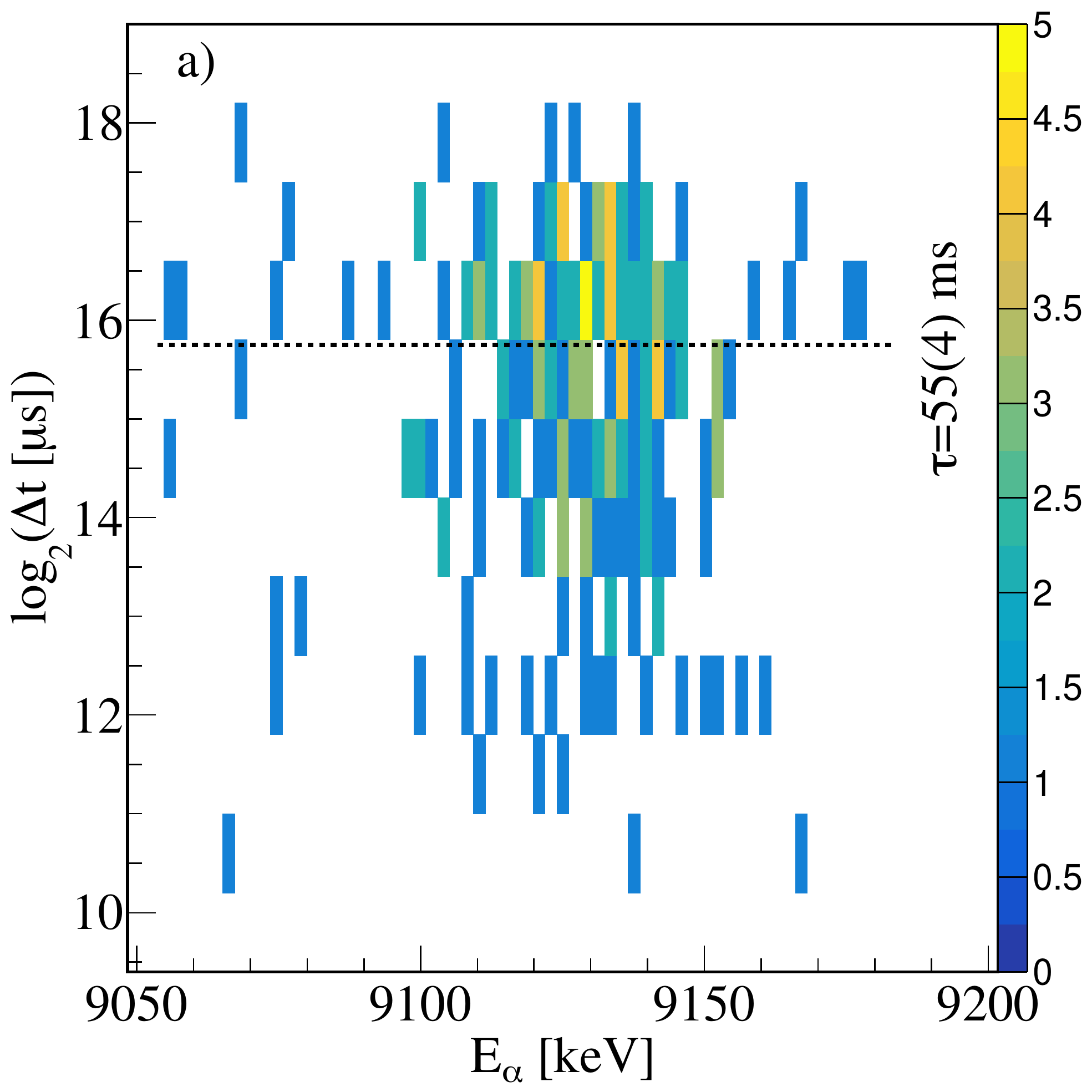}
  \includegraphics{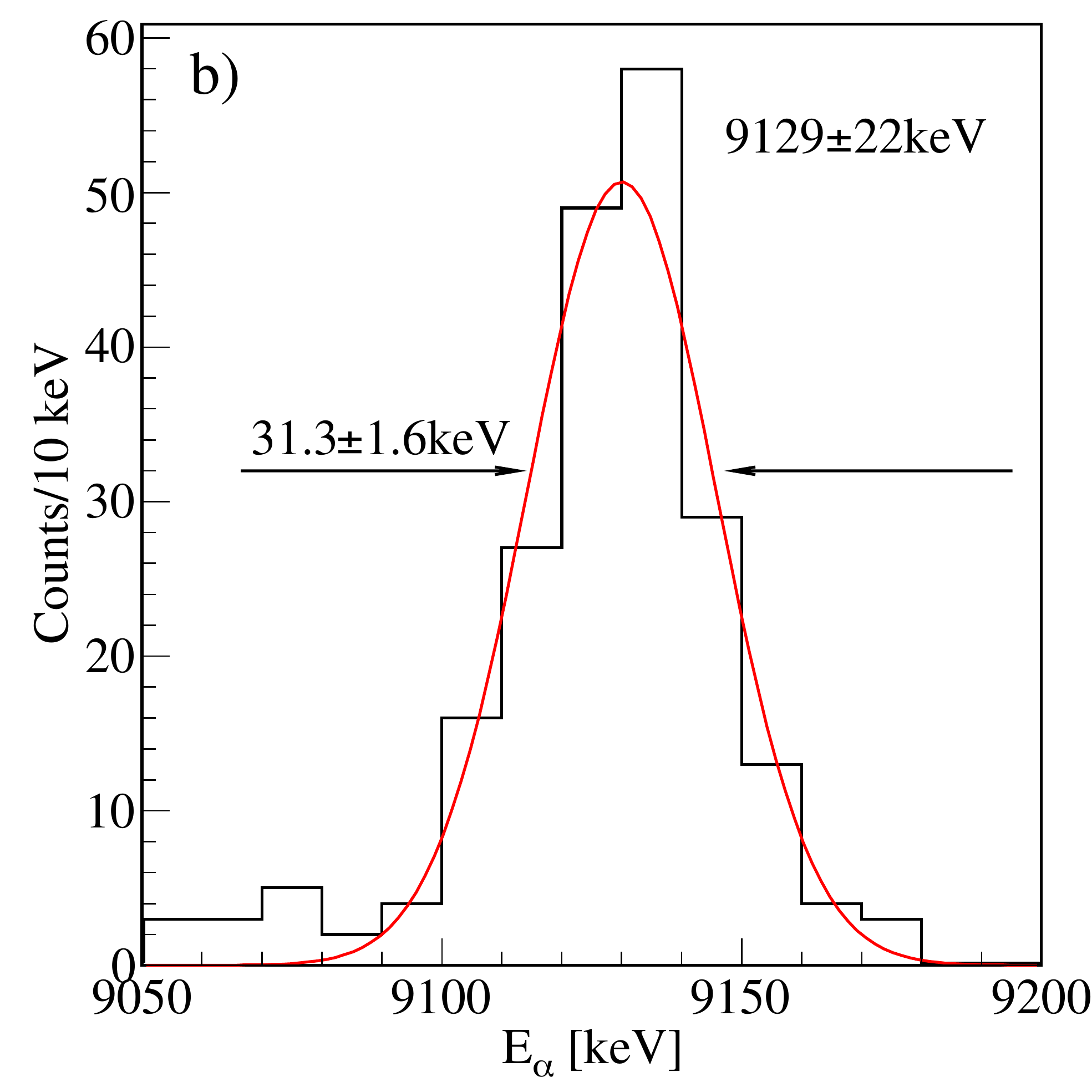}
  \includegraphics{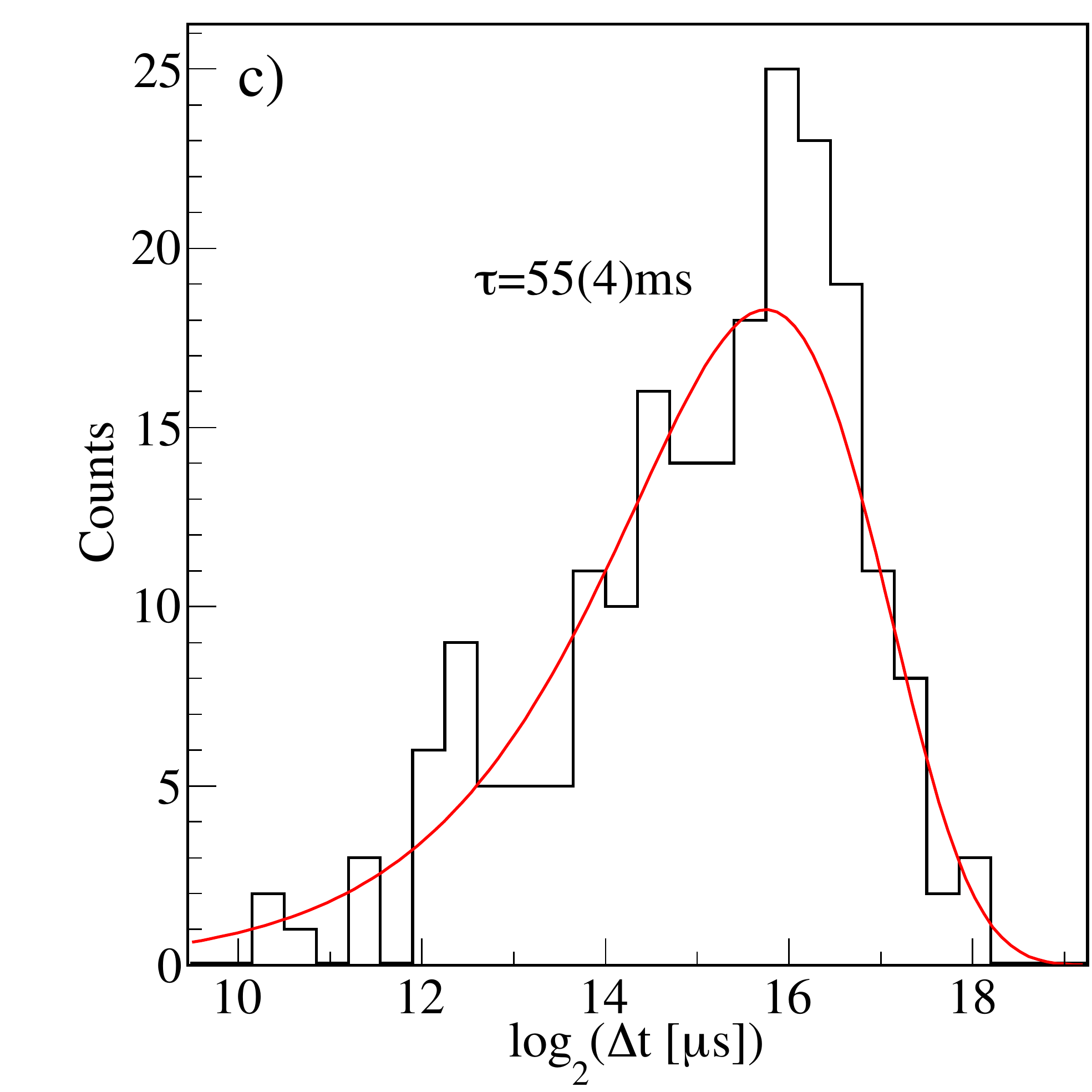}
}
\caption{a) The alpha-particles energy E$_{\alpha}$ of ER-$\alpha_{1}$ correlations as a logarithmic function of time difference $\Delta$t between implanted ERs and alpha decays detected in the same DSSD pixels. b) Energy projection of a) E$_{\alpha}$=9129(22) keV, FWHM=31.3(1.6) keV. c) Time projection of a) $\tau$=55(4) ms, T$_{1/2}$=38.3(2.8)ms }
\label{fig.249No}       
\end{center}
\end{figure*}
The investigation of the radioactive decay properties of transfermium elements is one of the main directions of modern nuclear physics and has recently made great  prog-ress due to the use of efficient detector arrays. Heavy and super heavy elements are mainly synthesized in fusion-evaporation reactions of heavy ions with heavy target nuclei with subsequent evaporation of several neutrons from the excited compound nucleus (CN). The Separator for Heavy Elements Spectroscopy (SHELS) \cite{Popeko,Yeremin} allows a separation of fusion-evaporation reaction products on the basis of velocities and charge states \cite{Eremin}. The identification of evaporation residues (ERs) is made through temporal and spatial correlations between registered decay events in the implantation detector of the GABRIELA detector array \cite{Hauschild,Chakma}.
The aim of the present experiment was to study the radioactive decay properties of nobelium isotopes produced in the fusion-evaporation reaction of an accelerated heavy ion beam of $^{48}$Ca with a $^{204}$PbS target. \\
At the FLNR, JINR, experiments directed at studying production cross-sections and radioactive decay properties of neutron-deficient nobelium isotopes have been previously successfully carried out. The first report about the spontaneously fissioning $^{250}$No nucleus can be found in Ref.\cite{Ter}, in which $^{250}$No was synthesized using the reaction $^{233}$U($^{22}$Ne, 5n) at the U-300 cyclotron. In 2003, Belozerov et al. used the reaction $^{204}$Pb($^{48}$Ca,xn) and measured twenty four long-lived and fifty six short-lived spontaneous fission (SF) events, which were assigned to $^{249}$No and $^{250}$No respectively \cite{Belozerov}. The measured half-lives of long-lived state attributed to $^{249}$No was  T$_{1/2}^{(SF)}$=$54.0_{-9.2}^{+13.9}$  $\mu$s and short-lived state attributed to $^{250}$No was T$_{1/2}^{(SF)}$=$5.6_{-0.7}^{+0.9}$ $\mu$s. The authors, however, did not exclude that both SF activities could be from $^{250}$No. In an experiment performed at the Fragment Mass Analyzer (FMA) in Argonne, it was shown that the previously-observed long-lived SF activity was in fact related to the decay of an isomeric state of $^{250}$No \cite{Peterson}. Subsequently, in an experiment using a 3He neutron multiplicity counter at the FLNR, the neutron multiplicities accompanying the fission of the ground and isomeric states were extracted \cite{Svirikhin}. An experiment performed at Jyväskylä measured an internal decay branch stemming from the isomeric state \cite{Kallu} evidencing a considerable fission hindrance due to K isomerism \cite{Xu}. More recently, Svirikhin et al. \cite{new} observed the decay of the new isotope $^{249}$No, which was produced with a large statistic. The same year, Ref. \cite{baatar} was published, where one decay event of $^{249}$No was observed. In this work, we present a detailed study of the data reported in Ref. \cite{new}. 
\section{Experimental details}
The recent study of neutron-deficient nobelium isotopes at the FLNR was performed using a $^{204}$PbS target. The 0.47(10) mg/cm${^2}$ target was made by electrodeposition on a 1.5 $\mu$m thick titanium foil backing. The enrichment of the target material was 99.94\% ($^{206}$Pb - 0.04\%; $^{207}$Pb - 0.01\%; $^{208}$Pb - 0.01\%). The $^{48}$Ca beam was delivered by the U-400 cyclotron with an average intensity of 0.5 p$\mu$A. The ERs produced in the fusion reaction of $^{48}$Ca+$^{204}$Pb were separated and transported through SHELS and delivered to the detection system GABRIELA. For the above-mentioned reaction, the transmission and detection efficiency of ERs was 20--40\% \cite{Yeremin} depending on the ion-optical settings of the separator. 

The GABRIELA system includes a 100x100 mm${^2}$ implantation Double-sided Silicon Strip Detector (DSSD) with 128$\times$128 horizontal and vertical strips. Upstream from the DSSD,  a Time of Flight detector (ToF) provides a marker to the DSSD events in order to distinguish recoil implants from subsequent decay events ($\alpha$ particles, electrons, SF fragments). The test reactions $^{48}$Ca+$^{174}$Yb and $^{48}$Ca+$^{164}$Dy were used for calibration purposes. In Fig. \ref{fig.calib} the spectrum of $\alpha$ particle energies detected in the DSSD is shown. The energy resolution of the DSSD is 15--20 keV for $\alpha$ particles ranging from 6--10 MeV. An additional silicon array consisting of eight DSSDs, each of 16$\times$16 strips, arranged in a tunnel configuration upstream from the implantation DSSD is used for internal conversion-electron spectroscopy and to increase the detection efficiency for $\alpha$ particles and SF fragments.  The detection efficiency of the implantation DSSD is 50\% for $\alpha$ particles and 100\% for fission fragments.

An array of germanium detectors is used for the detection of $\gamma$ rays and X rays. It is composed of four C-window coaxial single germanium crystals, arranged in a cross around the Si detectors and a clover detector installed behind the implantation DSSD. All the germanium detectors are surrounded by 15 mm thick BGO Compton-suppression shields in order to reduce the rate of background events and to improve the peak-to-total. The detection efficiency of $\gamma$ rays is $\sim$10\% at 600 keV and peaks at $\sim$30\% for photon energies around 100 keV \cite{Chakma}.
\section{Results}
The irradiation was performed at various bombarding energies. The details are given in Table \ref{tab.1}. 
\begin{table}
\caption{Details of target thicknesses \emph{d$_{t}$} used during the experiment as well as delivered beam energies \emph{E$_{lab}$} and doses. \emph{E$_{1/2}$} refers to the beam energy in mid-target}
\centering{}
\label{tab.1}      
\begin{tabular}{llll}
\hline\noalign{\smallskip}
\emph{E$_{lab}$} [MeV] & \emph{E$_{1/2}$} [MeV] & Projectile dose & d$_{t}$ [mg/cm$^{2}$]  \\
\noalign{\smallskip}\hline\noalign{\smallskip}
225.2 & 213.4 & 1.05$\times10^{18}$ &   \\
230.0 & 218.5 & 2.6$\times10^{17}$ &  \\
237.0 & 225.4 & 1.8$\times10^{18}$ & 0.47(10)  \\
242.0 & 231.0 & 6.3$\times10^{17}$ & \\
246.0 & 235.0 & 1.6$\times10^{17}$ &  \\
\noalign{\smallskip}\hline
\end{tabular}
\end{table}
\subsection{The isotope $^{249}$No and its daughter nuclei}
\begin{figure}[!ht]
\begin{center}
\resizebox{0.48\textwidth}{!}{%
  \includegraphics{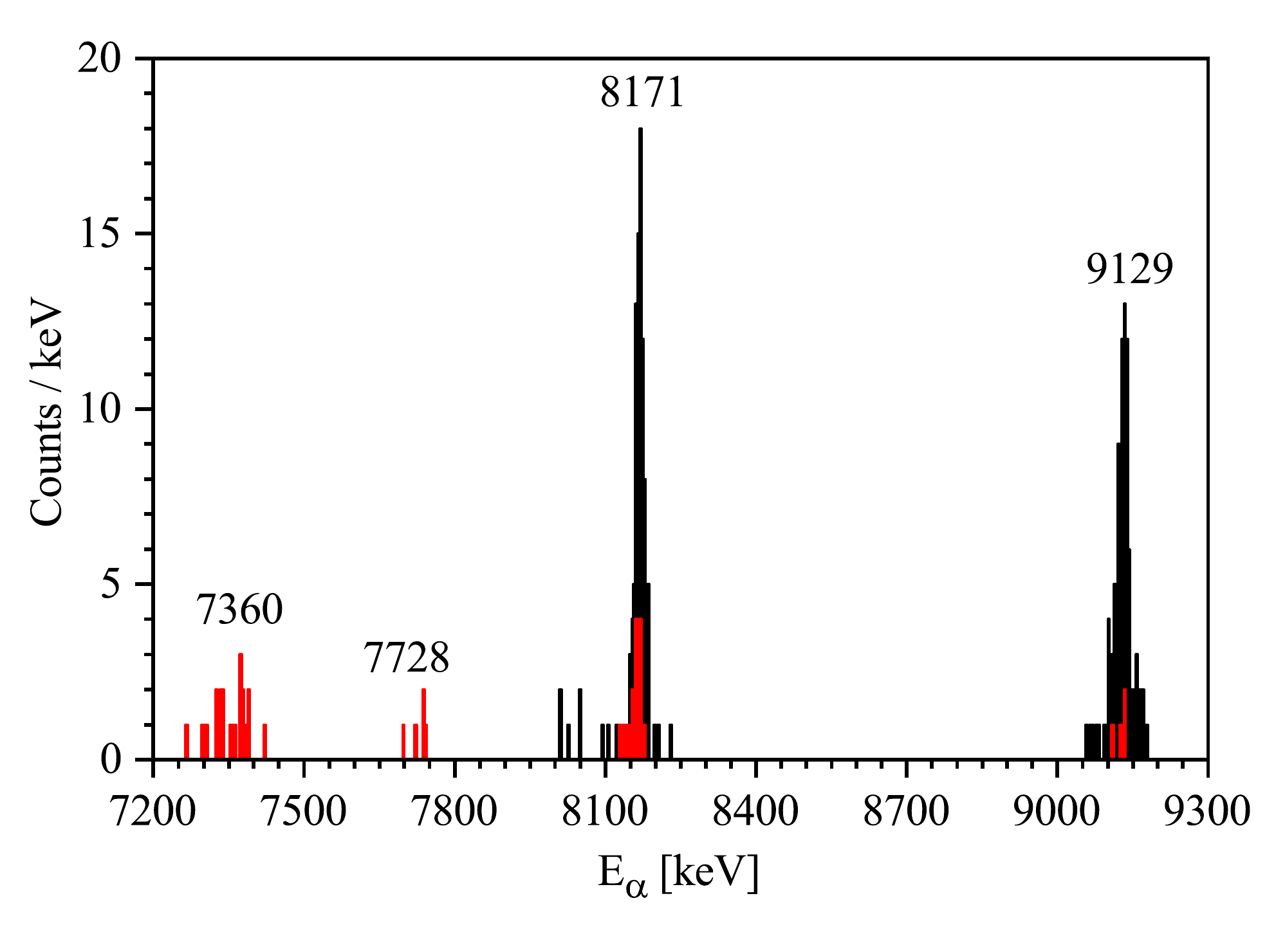}
}
\caption{Energy spectrum of alpha particles obtained from the genetic correlation analysis of the $^{249}$No alpha decay into known $^{245}$Fm (8171 keV) and $^{241}$Cf (7360 keV).The alpha activity with energy 7728 keV corresponds to the decay of $^{245}$Es. The black histogram corresponds to first and second generation decays following the implantation of an ER (ER-$\alpha_{1}$-$\alpha_{2}$).  The red histogram includes ER-$\alpha_{2}$-$\alpha_{3}$, ER-$\alpha_{1}$-$\alpha_{3}$ and ER-$\alpha_{1}$-$\alpha_{2}$-$\alpha_{3}$ correlations. See text for details}
\label{fig.spectr}       
\end{center}
\end{figure}

\begin{figure*}
\begin{center}
\resizebox{0.85\textwidth}{!}{%
  \includegraphics{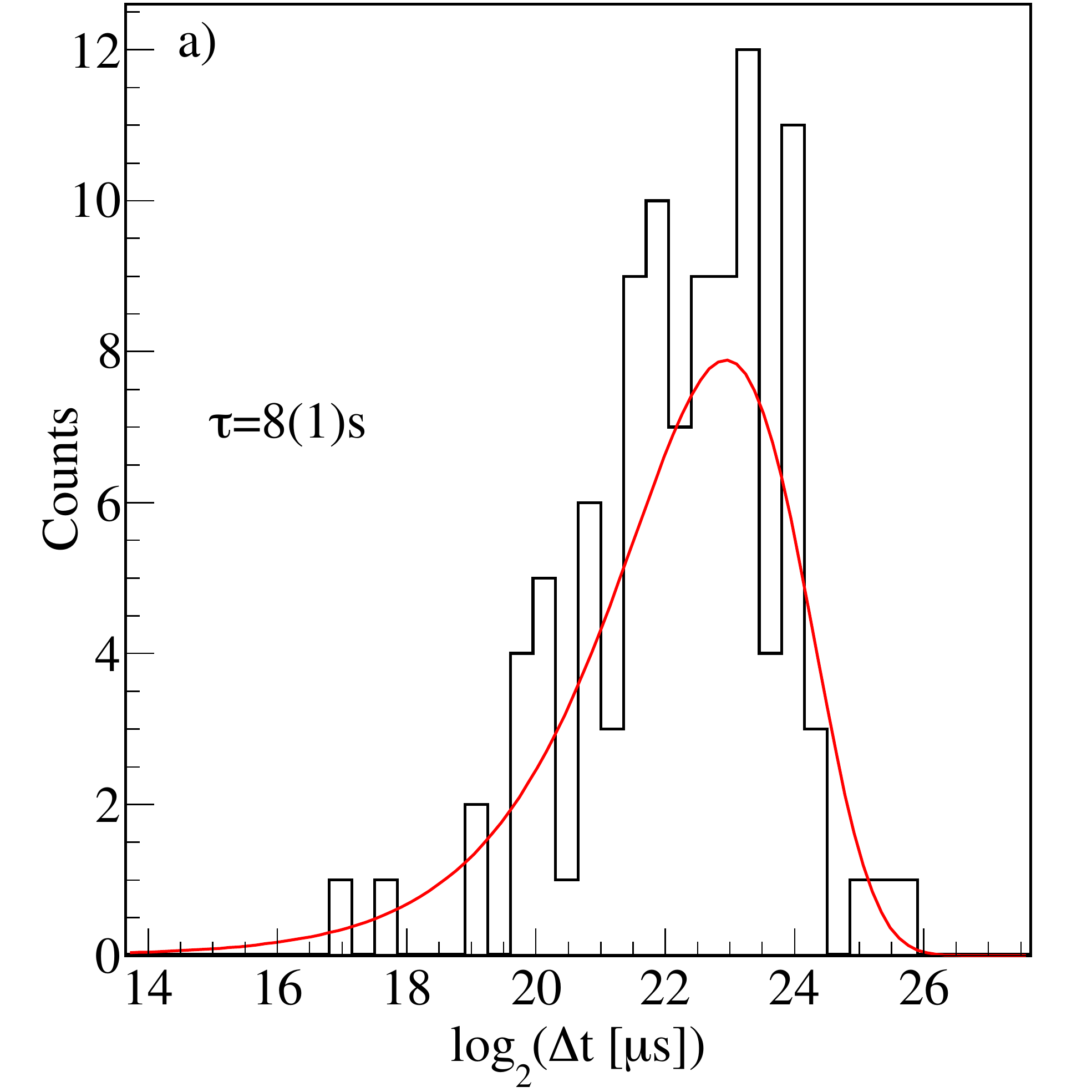}
  \includegraphics{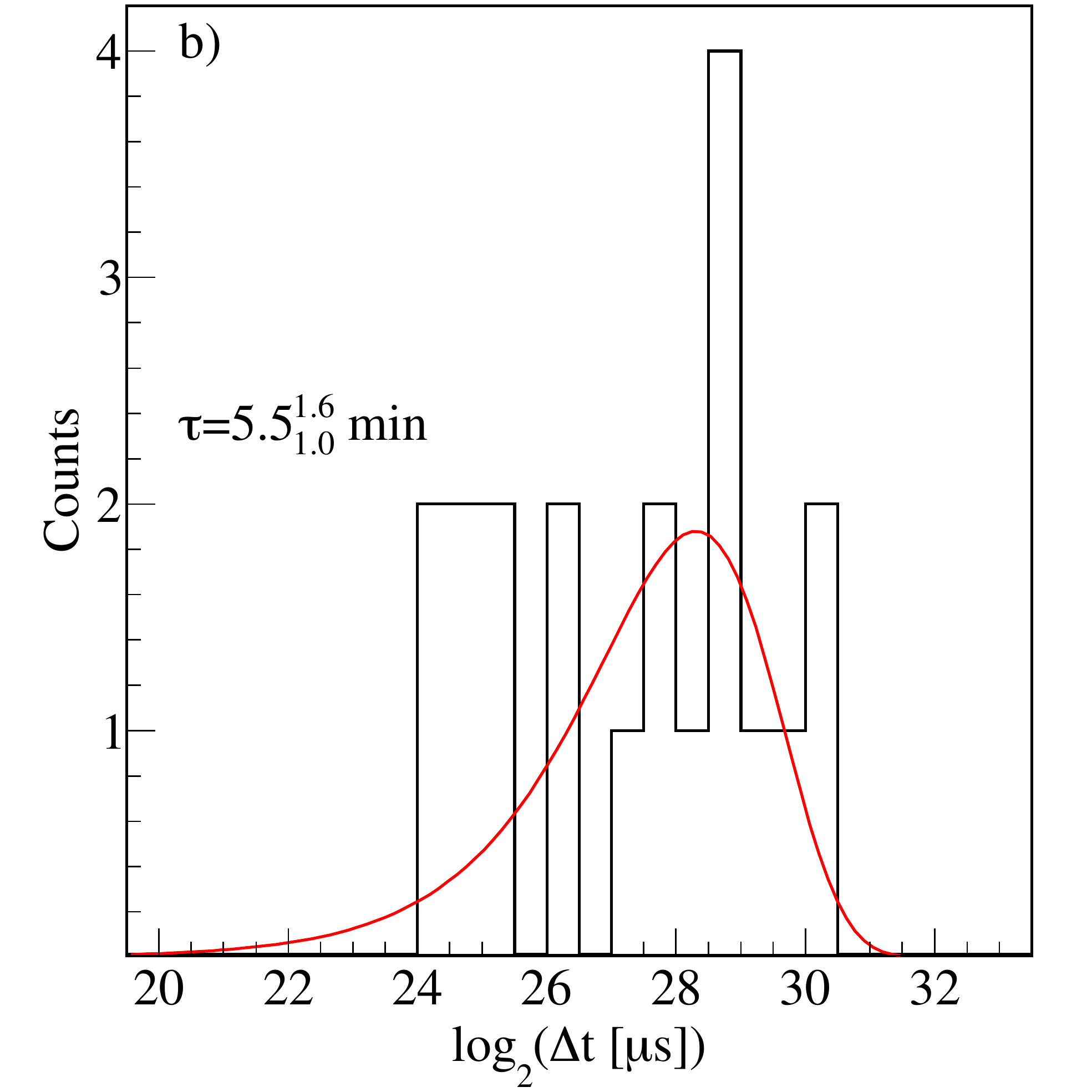}
  \includegraphics{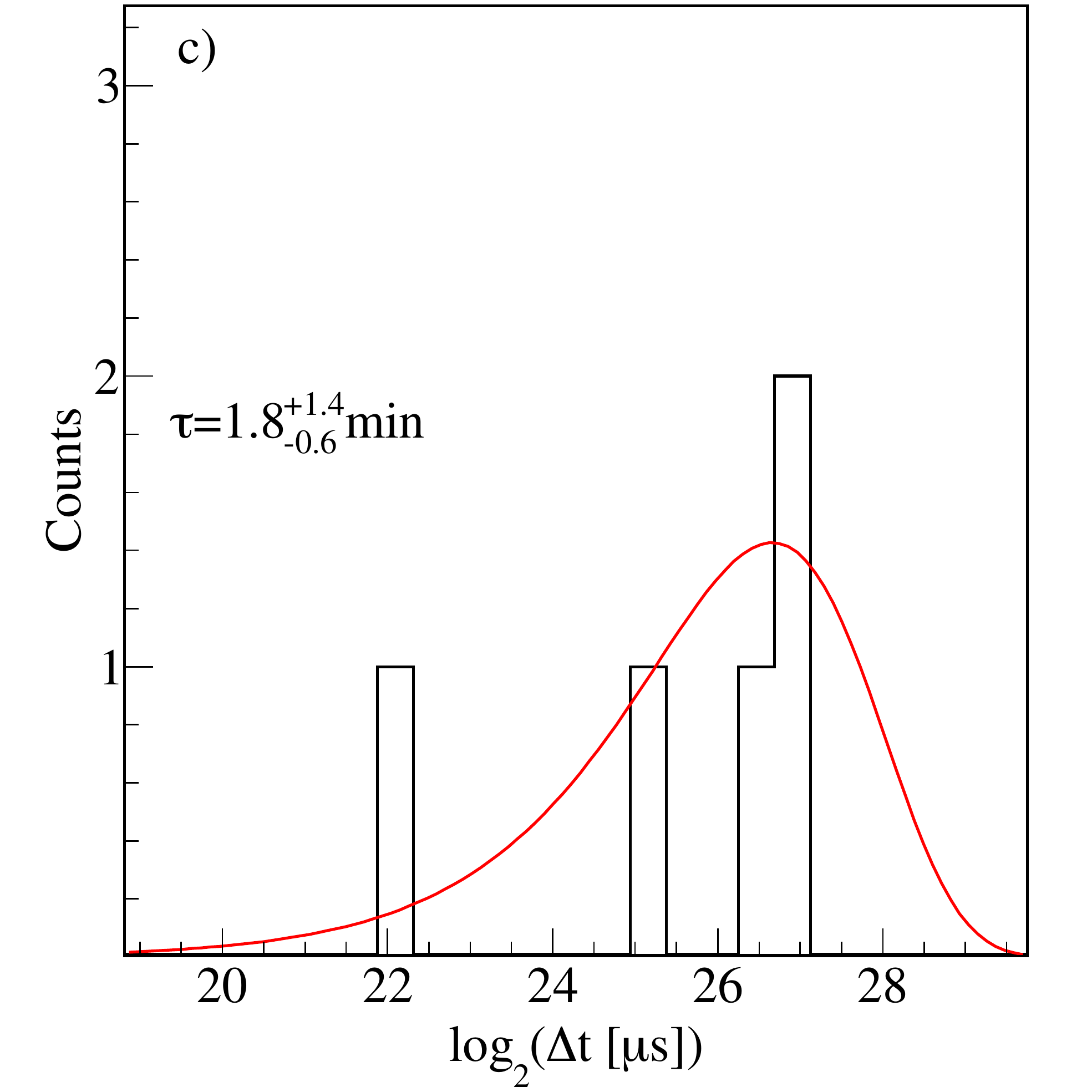}
}

\caption{Time distribution on a logarithmic time scale of the daughter events a) $^{245}$Fm with lifetime $\tau$=8(1)s, b) $^{241}$Cf, $\tau$=5.5$_{-1.0}^{+1.6}$ min  c) $^{245}$Es, $\tau$=1.8$_{-0.6}^{+1.4}$ min} 
\label{fig.timedecay}       
\end{center}
\end{figure*}

The isotope  $^{249}$No was produced in the fusion-evaporation reaction  $^{48}$Ca+ $^{204}$Pb$\rightarrow^{252}$No$^{*}\rightarrow^{249}$No+3n. We searched for alpha decays in the energy range 8-10 MeV within a time window of 500 ms after the implantation of ERs with energies 1-18 MeV. Altogether 218 ER-$\alpha$ correlation events in the range of 9050--9200 keV (Fig. \ref{fig.249No}a) were observed. This new activity has a mean energy of E$_{\alpha}$=9129(22) keV. From a fit of the time distribution of Fig. \ref{fig.249No}c, the lifetime is found to be to $\tau$=55(4) ms, corresponding to a half-life 38.3(2.8) ms. The number of observed ER-$\alpha$ correlations as a function of beam energy is given in Table \ref{tab.2}. 
\begin{table}
\caption {Production yields of ER-$\alpha$ correlations (N$_{ER-\alpha}$) for the new isotope $^{249}$No in the reaction $^{204}$Pb($^{48}$Ca, 3n)$^{249}$No. The cross-sections  $\sigma$(3n) were calculated assuming a 34\% transmission efficiency of SHELS. For the point at 225.4 MeV a statistical uncertainty has been computed, for other points of beam energies uncertainties have been calculated according to the prescriptions of Ref. \cite{Schmidt}. The uncertainty of transmission was not taken into account.  The quoted beam energies \emph{E$_{1/2}$} are mid-target energies }
\centering{}
\label{tab.2}       
\begin{tabular}{llll}
\hline\noalign{\smallskip}
\emph{E$_{1/2}$} [MeV]   & N$_{ER-\alpha}$   &  $\sigma$(3n) [nb]  \\
\noalign{\smallskip}\hline\noalign{\smallskip}
 218.5 & 2  & 0.03$_{-0.02}^{+0.04}$   \\
225.4 & 193 & 0.47(4)  \\
231.0 & 22 & 0.15$_{-0.03}^{+0.04}$  \\
235.0 & 1 & 0.03$_{-0.04}^{+0.07}$\\
\noalign{\smallskip}\hline
\end{tabular}
\end{table}
A genetic correlation analysis (shown in Fig. \ref{fig.spectr} and Table \ref{tab.3}) reveals that the newly-discovered activity correspond to the alpha decay of the new isotope $^{249}$No as it is followed by the characteristic alpha decay of $^{245}$Fm and subsequently $^{241}$Cf. Indeed, the alpha particle energies of 8171(20) keV and 7360(27) keV with corresponding half-lives of 5.5(7) s and 3.8$_{-0.7}^{+1.1}$ min (see Fig. \ref{fig.timedecay} a and b) are in agreement with the tabulated values for $^{245}$Fm and $^{241}$Cf respectively \cite{Kondev,Nurmia,Silva}.  
\begin{figure}[!ht]
\begin{center}
\resizebox{0.35\textwidth}{!}{%
  \includegraphics{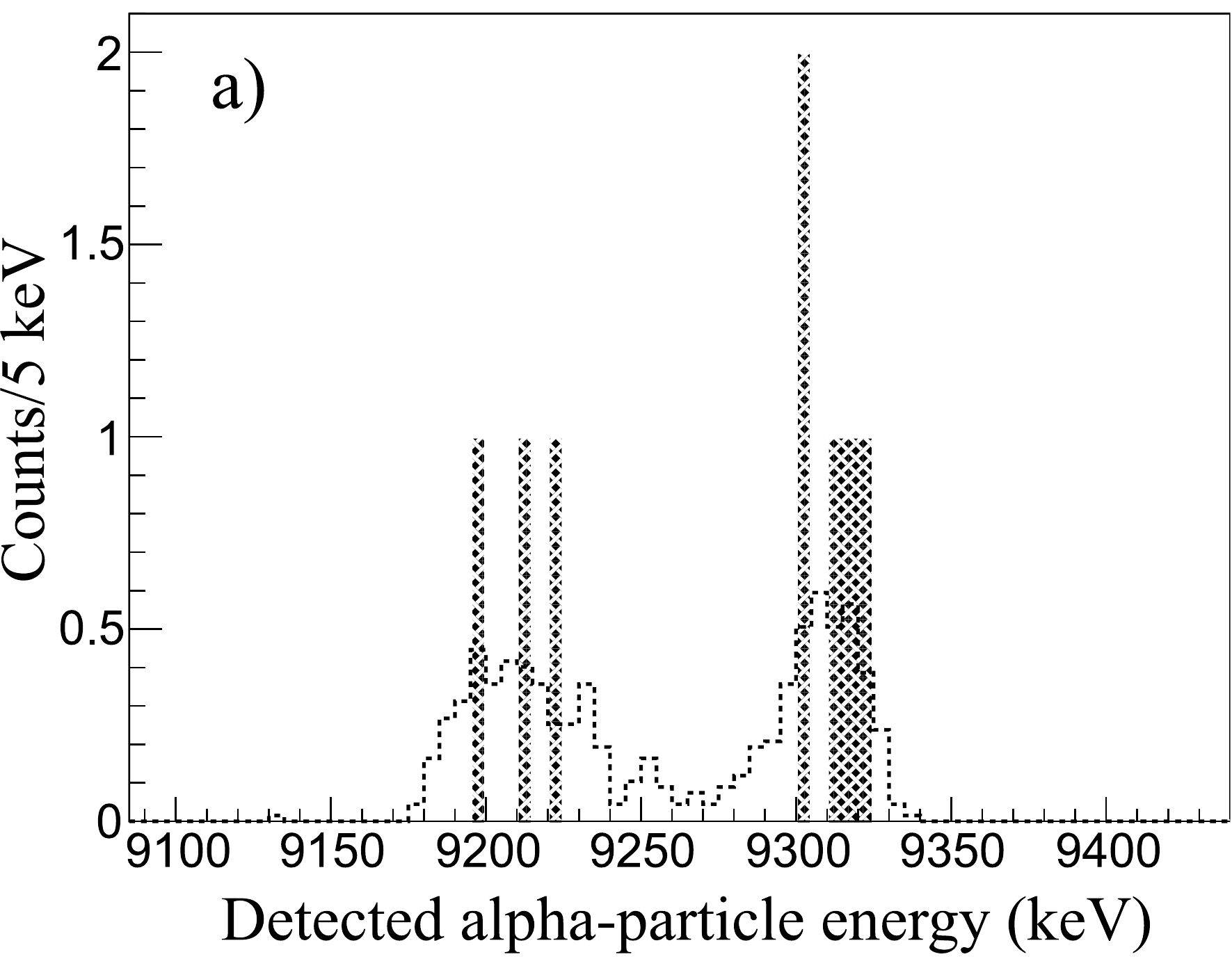}
}
\resizebox{0.36\textwidth}{!}{%
  \includegraphics{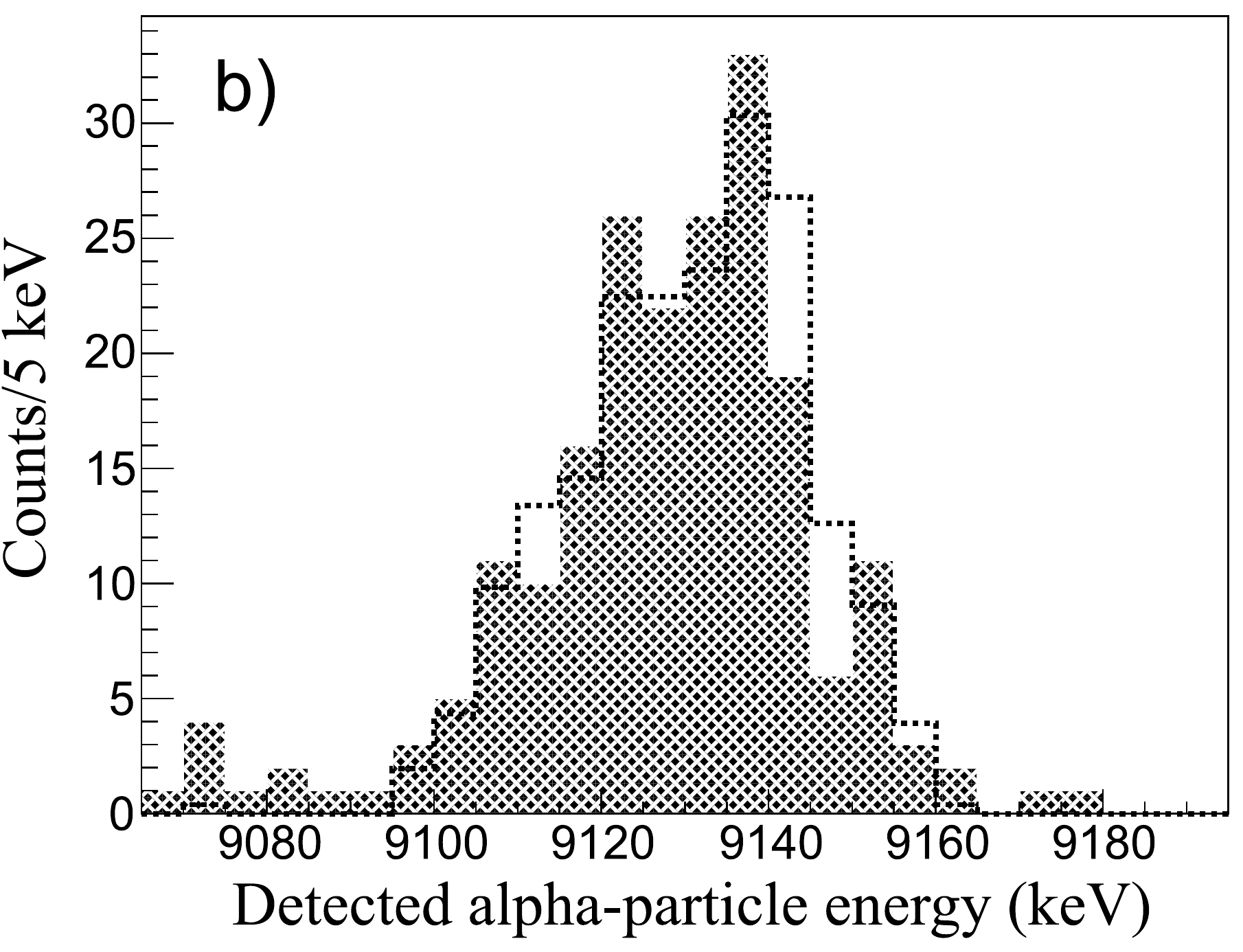}
}
\caption{a) Spectrum of $\alpha$ decay energies registered in the case of the $\alpha$ decay of $^{253}$Rf (shaded histogram) and simulated spectrum (dashed histogram) assuming a 95$\%$ branch to a 1/2$^{+}$ state at $\sim$125 keV excitation energy, followed by an E2 decay to a 5/2$^+$ ground state and normalised to the experimental number of counts.  b) Same as a) in the case of the decay of $^{249}$No and assuming that the decay proceeds through a 5/2$^+$ state at $\sim$50 keV above the ground state, which decays by emission of a $\sim$30$~$keV M1 transition to the 3/2$^+$ member of the ground state band followed by a  $\sim$10$~$keV M1 transition to the ground state.}
\label{fig.sim}       
\end{center}
\end{figure}
Only one ER-SF event with a decay time 344 ms was observed at the beam energy 225.4 MeV. This event cannot be unambiguously assigned to $^{249}$No since this SF event also may belong to $^{251}$No or $^{252}$No, which was produced in the 2n channel on the $^{206}$Pb impurities of the target. Indeed, about 10 ER-$\alpha_{1}$($^{252}$No)-$\alpha_{2}$($^{248}$Fm) correlated events were detected. The alpha decay of to $^{254}$No was also observed due to the presence of to $^{208}$Pb impurities and the rather large cross-section of the reaction $^{208}$Pb($^{48}$Ca,2n) $^{254}$No. Therefore an upper limit for the SF branch (for one event) of $^{249}$No is b$_{SF}$$\leq$(2.3$_{-2.3}^{+4.6})\times10^{-3}$. 
\begin{table}[th]
\begin{center}
\caption{Number of detected evaporation residue events with subsequent alpha particles obtained for the different correlation combinations for the decay of $^{249}$No in the reaction $^{204}$Pb($^{48}$Ca, 3n)$^{249}$No}
\label{tab.3}       
\begin{tabular}{l|lllll}
\hline\noalign{\smallskip}
Type of correlations &   Number of events \\
\noalign{\smallskip}\hline\noalign{\smallskip}
ER-$\alpha_{1}$ & 218 \\
ER-$\alpha_{1}$-$\alpha_{2}$ & 101 \\
ER-$\alpha_{1}$-$\alpha_{2}$-$\alpha_{3}$ & 1 \\
ER-$\alpha_{1}$-$\alpha_{3} $ & 4 \\
ER-$\alpha_{2}$-$\alpha_{3}$ & 16 \\
\noalign{\smallskip}\hline
\end{tabular}
\end{center}
\end{table}

The FWHM of the $^{249}$No alpha-particle energy peak ($\sim$30keV) is larger than the one of the $^{245}$Fm peak ($\sim$20keV). This indicates that the alpha decay of $^{249}$No populates low-lying excited states in the daughter nucleus. This is at variance with the broad alpha-decay spectrum from the 7/2$^+$ ground state of the isotone $^{247}$Fm \cite{Hessberg}.  Another difference between the 2 isotones is the fact that only one alpha-decaying state is observed in $^{249}$No. The different low-lying structure of $^{249}$No as compared to $^{247}$Fm is also revealed by the decay pattern of $^{253}$Rf \cite{LopezRf} (see panel a) of Fig. \ref{fig.sim}). The alpha decaying state in $^{253}$Rf has been identified as the low-spin 1/2$^{+}$[631] neutron state and the corresponding alpha decay spectrum shows 2 emissions at 9.21 and 9.31 MeV, both followed by the characteristic $\alpha$ decay of $^{249}$No.
\begin{figure*}[h!]
\begin{center}
\resizebox{0.65\textwidth}{!}{%
  \includegraphics{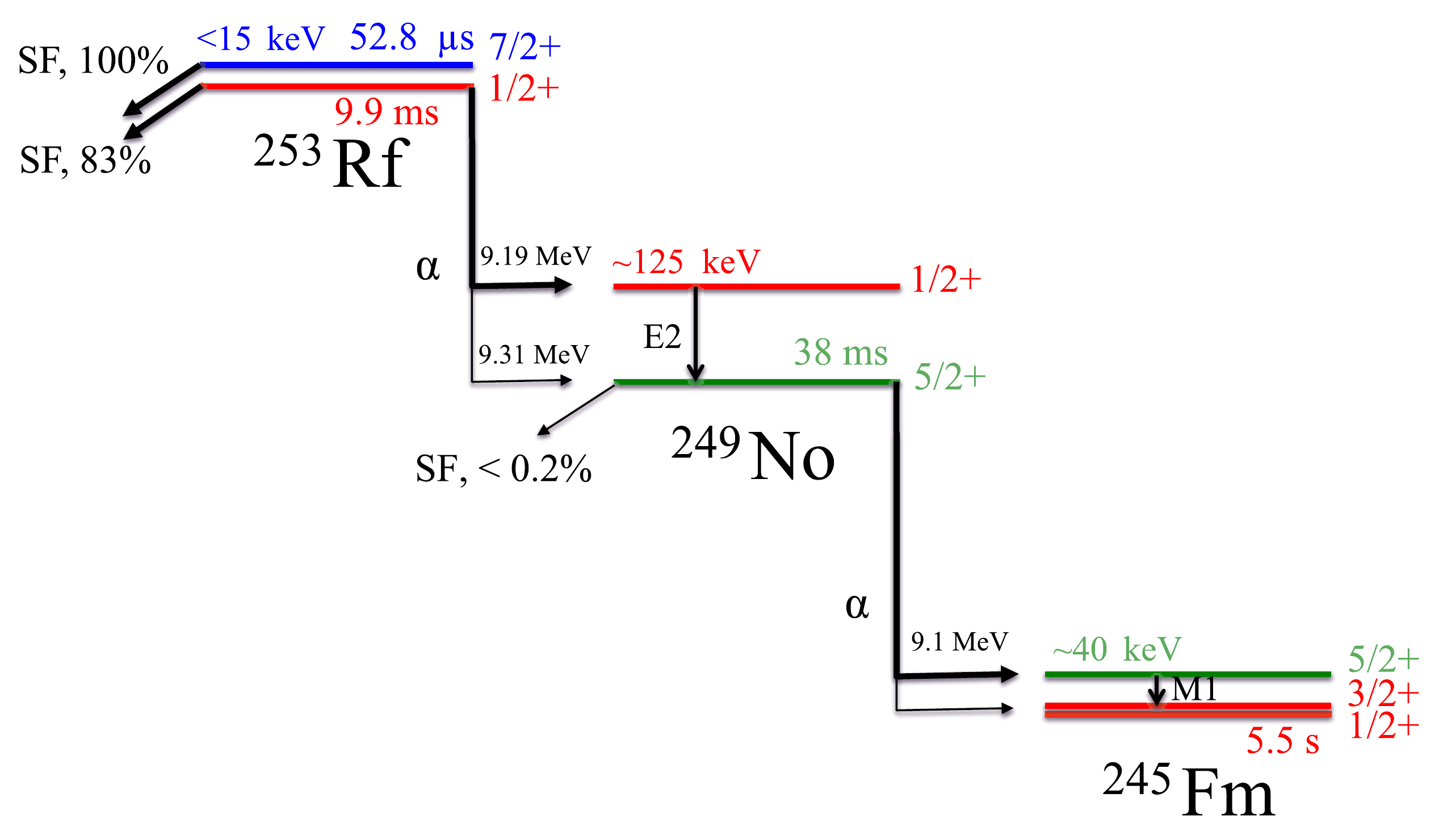}
 }
\caption{Decay scheme of $^{253}$Rf and $^{249}$No derived from the present experiment and data from \cite{LopezRf}.}
\label{fig.scheme}       
\end{center}
\end{figure*}
\begin{figure}[h!]
\begin{center}
\resizebox{0.45\textwidth}{!}{%
  \includegraphics{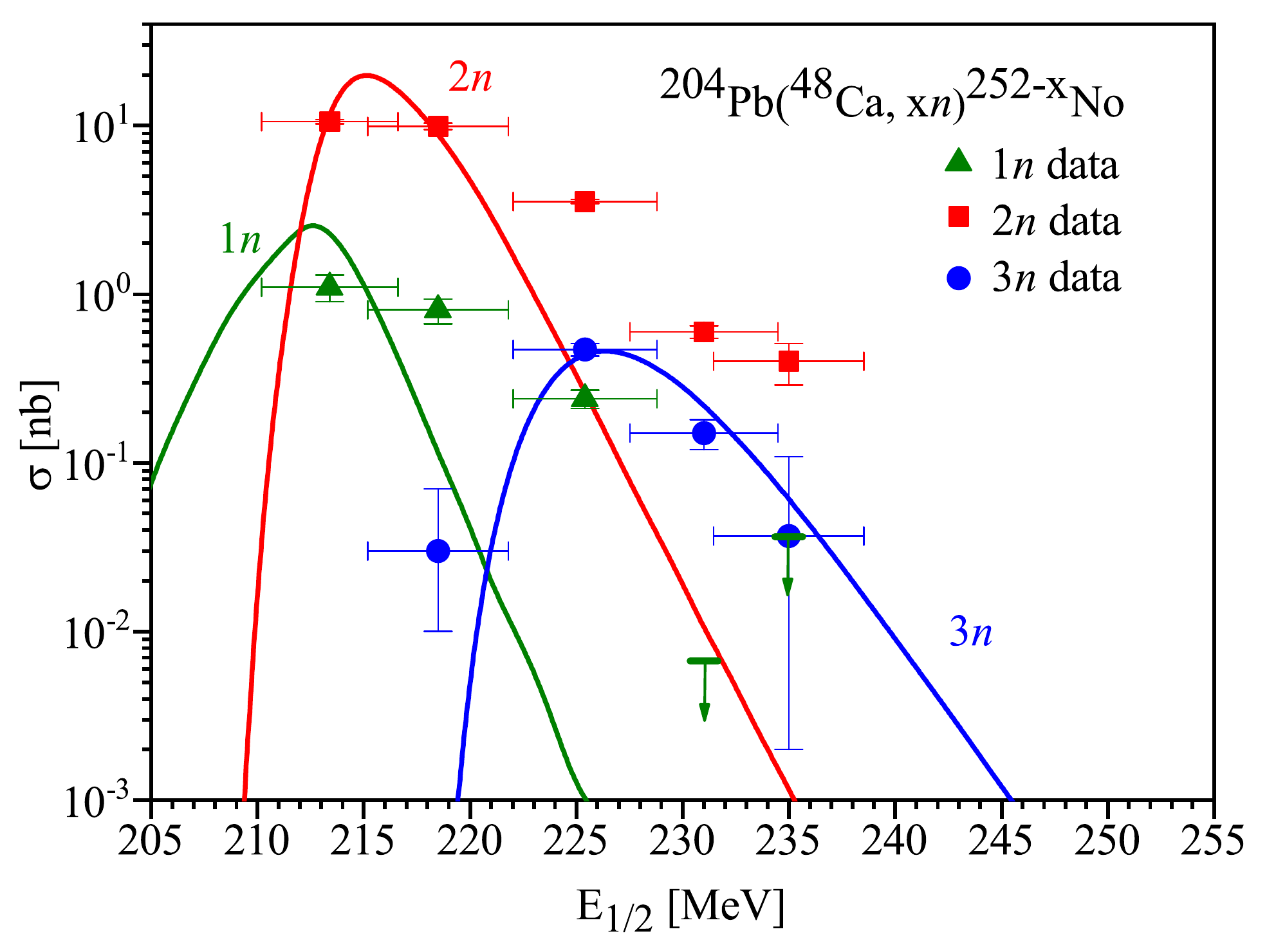}
}
\caption{Production cross-sections of nobelium isotopes in the xn-channels of the reaction $^{48}$Ca+$^{204}$Pb$\rightarrow^{252}$No$^{*}$ as a function of the mid-target beam energy E$_{1/2}$. The results of the NRV calculations are represented by solid lines \cite{Karpov,AVKarpov}. The measured cross-sections for the 1n channel ($^{251}$No) are denoted by green solid triangles and for the last two energies (231 and 235 MeV) cross-section limits were pointed. The solid red squares represent the cross-sections of the 2n channel ($^{250}$No). The evaluated cross-section of the 3n channel (new isotope $^{249}$No) are marked by blue circles. On the x-axis error bars represent a measurement variability of the beam energy, on the y-axis error bars are statistical \cite{Schmidt}}
\label{fig.cross}       
\end{center}
\end{figure}

These features can be explained as arising from a favou-red alpha decay to the 1/2$^+$ state, which subsequently decays by a $\approx$125 keV converted E2 transition to a 5/2$^+$ ground state. This decay scenario is supported by the results of Geant4 simulations, performed as described in Ref. \cite{Chakma} . The simulated distribution of detected $\alpha$-particle energies is shown together with the experimental one in panel a) of  Fig. \ref{fig.sim}. Despite the obvious lack of statistics in the experimental spectrum, its overall properties are well reproduced by a $\approx$95$\%$ alpha branch to the 1/2$^+$ state. The ground state of $^{249}$No is therefore assigned the 5/2$^+$[622] neutron configuration, as in the lighter isotones $^{239}$U, $^{241}$Pu and $^{243}$Cm. 

This assignment also makes sense if one considers the fission properties of the ground state of $^{249}$No. If the ground state were based on the 7/2$^+$[624] configuration like $^{247}$Fm, its fission hindrance should be much smaller than the 1.02$\times$10$^{5}$ fission hindrance of the 7/2$^{+}$ ground state of the heavier isotope $^{251}$No (see Ref. \cite{Hessber} for details). This is not the case since the fission hindrance of $^{249}$No with respect to its only known even-even neighbour $^{250}$No is extracted to be at least 40 times larger. The difference in fission hindrance is therefore due to a different single particle structure and the 5/2$^+$[622] configuration must be associated with a much larger specialisation energy than the 7/2$^+$[624] configuration. Finally, if one assumes that the ground state of $^{245}$Fm is based on the 1/2$^+$[631] neutron orbital (from systematics of the N=145 isotones), the broad alpha peak of $^{249}$No can be explained if the 5/2$^+$ state in $^{245}$Fm lies $\approx$40 keV above the ground state. 

Geant4 simulations show that a cascade composed of a $\approx$30 keV M1 transition to the 3/2$^+$ member of the ground state band followed by a small $\approx$10 keV intraband M1 transition to the ground state can account for the features of the experimental spectrum (see panel b) of Fig. \ref{fig.sim}). The 1/2$^+$ state cannot lie much higher than $\approx$50 keV above the ground state as then summing of the $\alpha$-particle energy and the particles emitted in the internal-conversion process of the $>$40 keV M1 transition would lead to a broader apparent peak with a low-energy shoulder. This maximal energy depends however on the exact spacing between the 3/2$^+$ member of the ground state band and the ground state, which is known to vary between 5--12 keV in the lighter N=145 and 147 isotones. The proposed decay scheme of $^{253}$Rf through $^{249}$No  to $^{245}$Fm is shown in Fig. \ref{fig.scheme}.  The uncertainty in the $^{249}$No decay scheme leads to an $\alpha$-decay energy Q$_{\alpha}$ of 9.28(3) MeV, which is consistent  to the value extracted from the mass evaluation table 9170(200) keV \cite{Wang}.

A group of alpha particles with an energy of 7728(20) keV and half-life 1.2$_{-0.4}^{+1.0}$ min (see Fig. \ref{fig.spectr}c) was also found to follow the decay of $^{249}$No. This activity corresponds to the alpha decay of $^{245}$Es, which is produced by $\beta^{+}$ or the electron capture (EC)  of $^{245}$Fm. From the known alpha-decay branch of $^{245}$Es, the EC or $\beta^{+}$ branch of $^{245}$Fm was extracted to be b$_{EC/\beta^{+}}$=(11.5$_{-5.0}^{+6.8}$)\%. In an experiment performed at GSI aimed at studying the $^{245}$Fm isotope \cite{Khuyagbaatar}, the authors suggested that the observation of $^{245}$Es decays may be due to direct production of $^{245}$Es via the p2n channel of the $^{40}$Ar+$^{208}$Pb$\rightarrow^{248}$Fm$^{*}$ reaction.  At the same time, they did not exclude a contribution of a possible EC branch of $^{245}$Fm and deduced an upper limit of b$_{EC}$= 7\%, which is consistent with our measured value. 

The measured excitation function of $^{249}$No is shown in Fig. \ref{fig.cross} by the solid blue circles together with theoretical calculations performed using NRV \cite{Karpov,AVKarpov} within the statistical model. The de-excitation of the CN was simulated using the method of nested integrals, which is capable of allowing for channels with evaporation of a limited number of particles \cite{Kar}.  The experimental data follow the shape of the theoretical curve for the 3n channel. A production cross-section value of 0.47(4) nb was obtained from the number of detected $\alpha$-decay events at the beam energy E$_{1/2}$=225.4 MeV that correspond to the maximum of an excitation function.  
\begin{figure*}[tb]
\begin{center}
\resizebox{0.85\textwidth}{!}{%
  \includegraphics{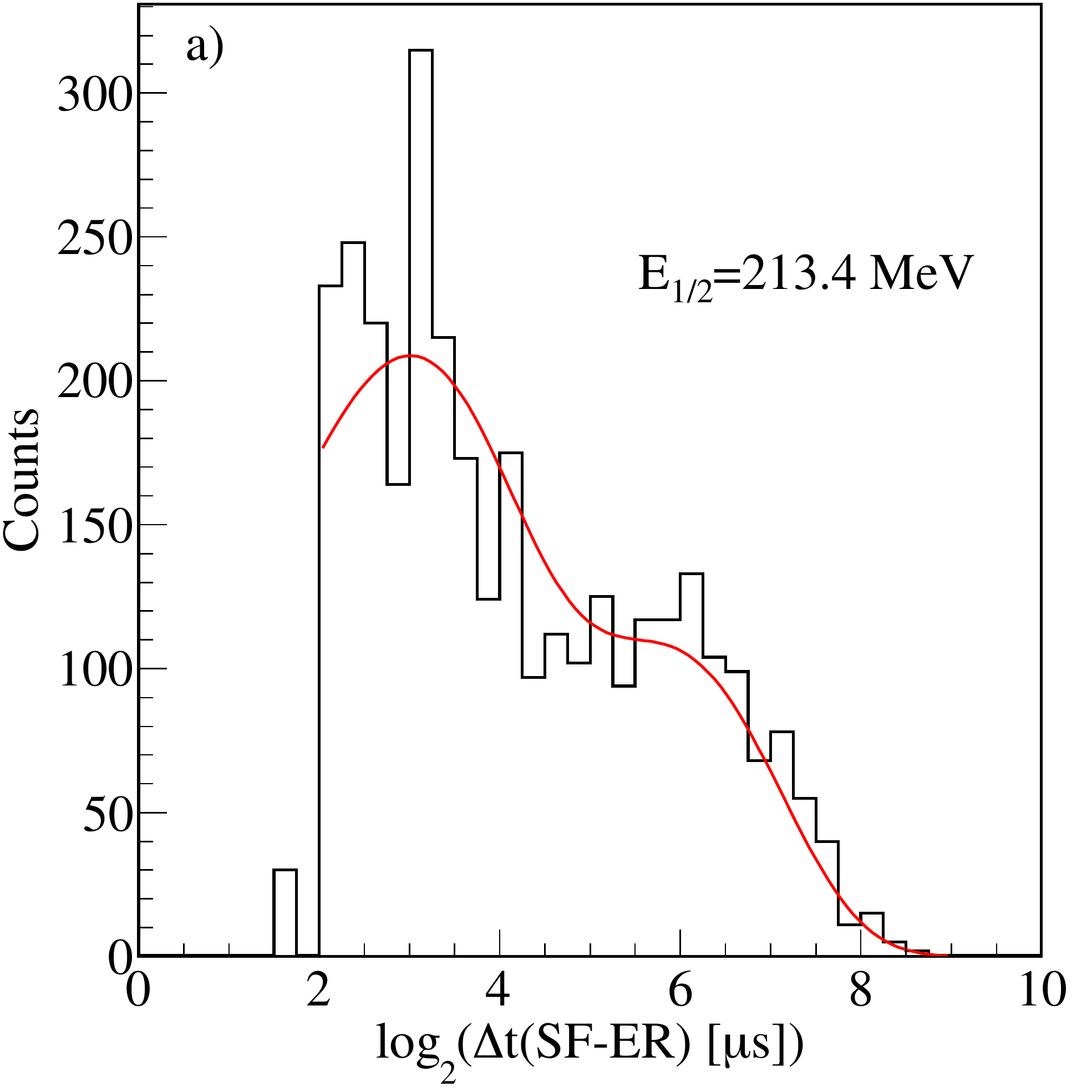}
  \includegraphics{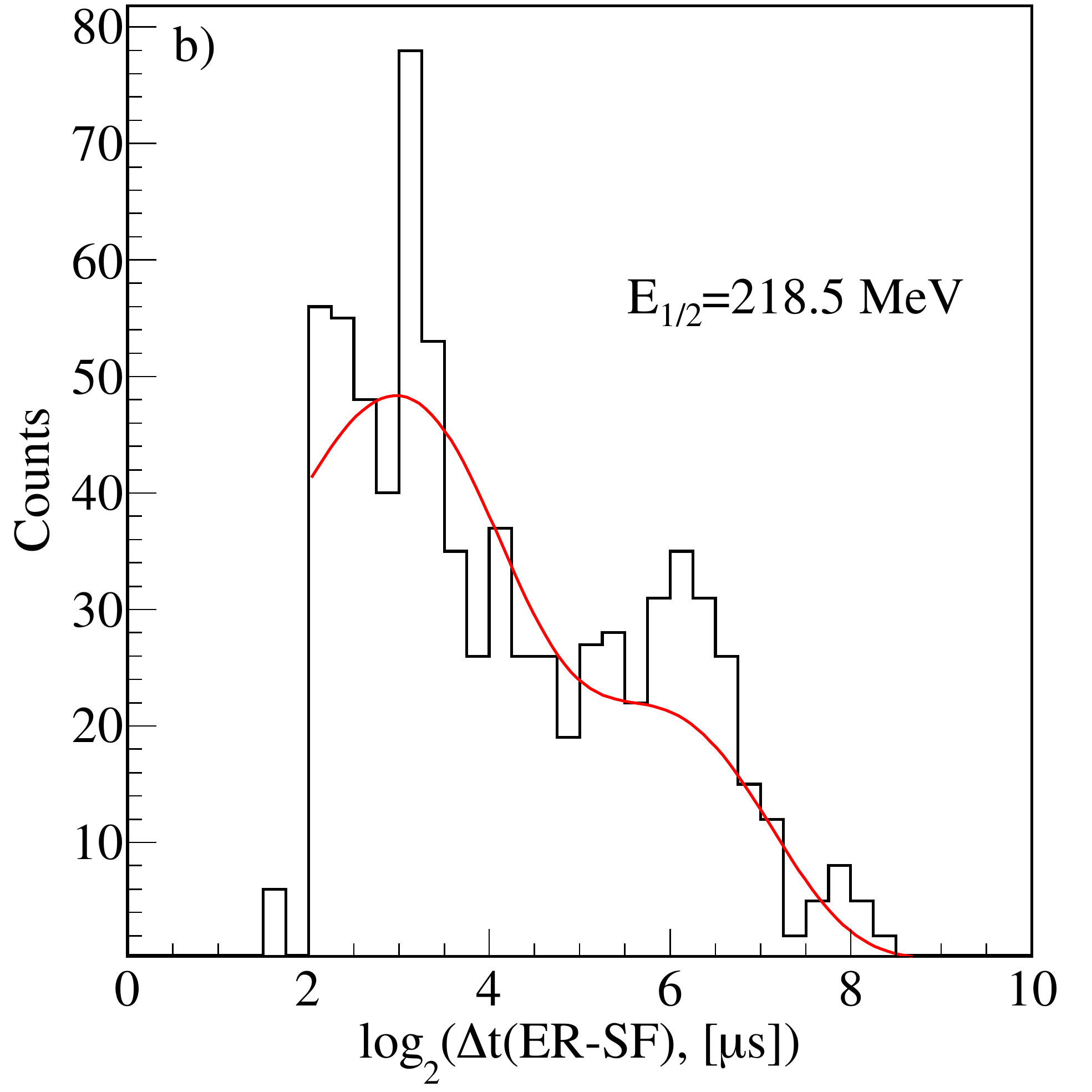}
  \includegraphics{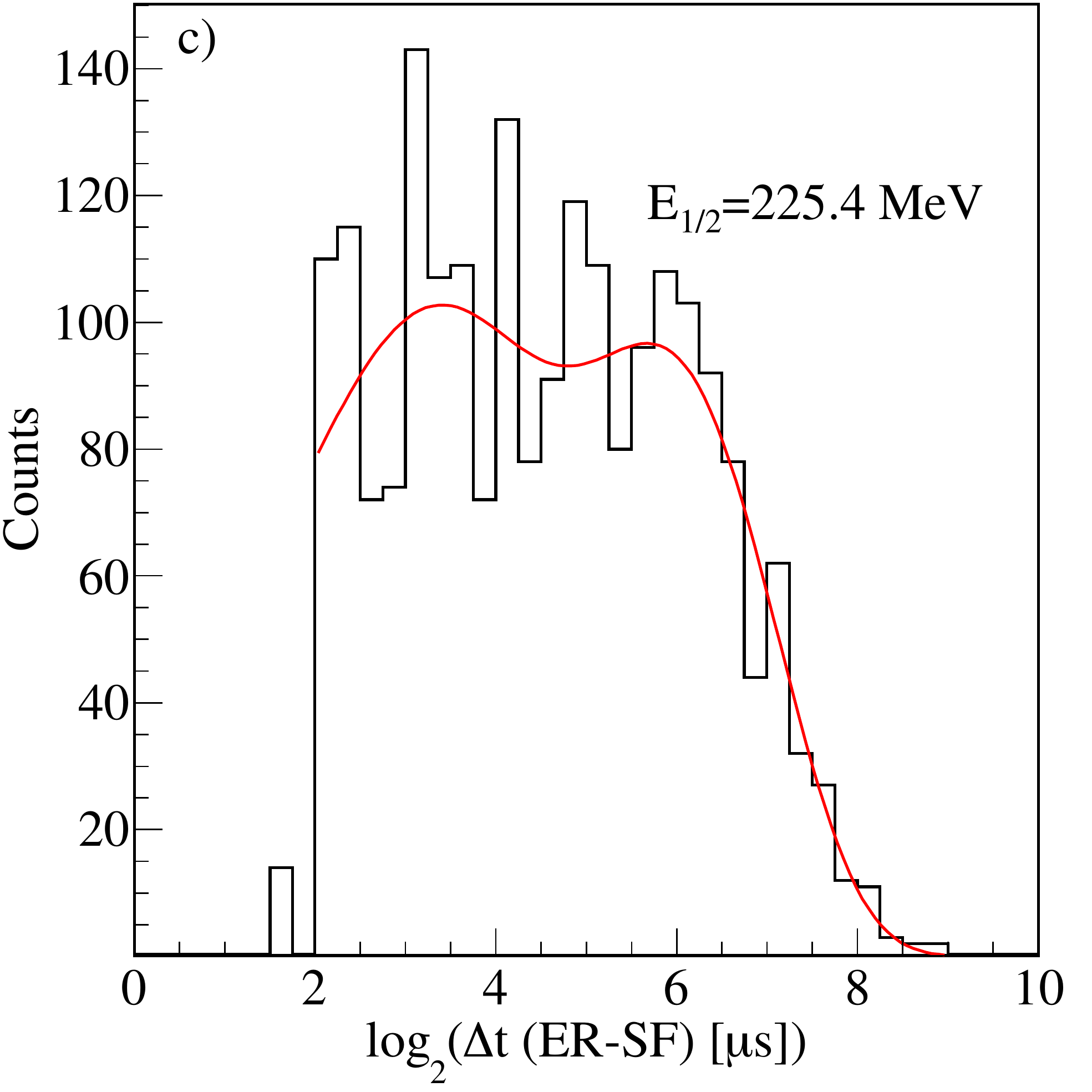}
}
\caption{Time distribution of correlated ER-SF events for the ground and isomeric states of $^{250}$No at different beam energies: a)213.4 MeV, b)218.5 MeV and c)225.4 MeV. The solid red lines are the fitted two-component exponential decay curves}
\label{fig.decaytime}    
\end{center}
\end{figure*}
\subsection{The isotopes $^{250}$No and $^{251}$No}
\begin{figure}[h!]
\begin{center}
\resizebox{0.48\textwidth}{!}{%
  \includegraphics{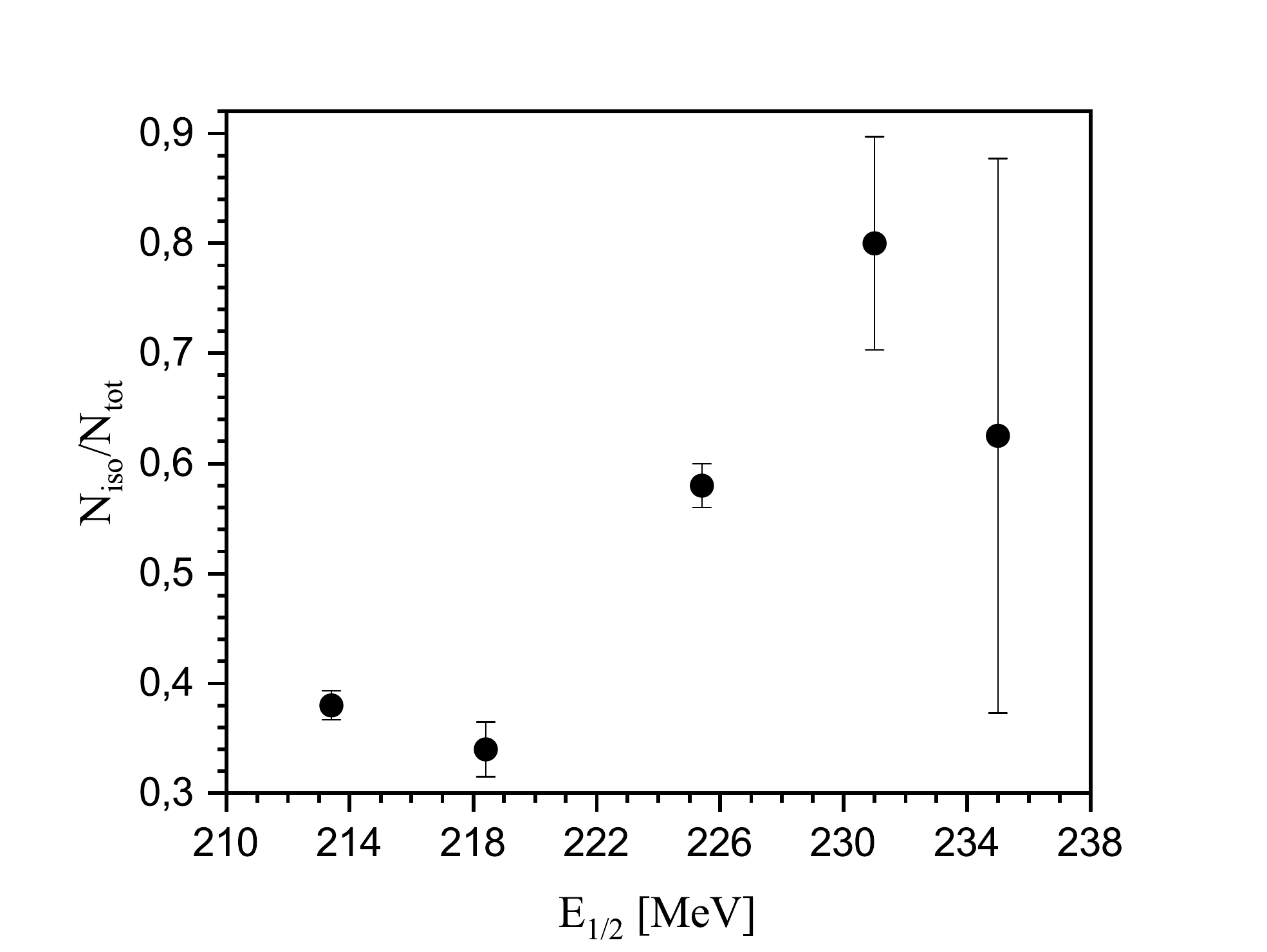}
}
\caption{$^{250}$No levels population N$_{iso}$/N$_{tot}$ change depending on beam energies (E$_{1/2}$). N$_{iso}$ is the longer-lived isomeric state events, N$_{tot}$ is the total SF events of $^{250}$No during all irradiations }

\label{fig.ratio}       
\end{center}
\end{figure}

The isotope $^{250}$No was produced in the 2n channel of the reaction $^{48}$Ca+$^{204}$Pb. A total production cross-section for the 2n evaporation channel was extracted considering the total number of detected ERs followed by spontaneous fission in the same pixel of the DSSD (solid red squares in Fig. \ref{fig.cross}). The ER-SF correlations for $^{250}$No represent events of the short-lived ground state T$_{1/2}$=4.7(1) $\mu$s and long-lived high-K isomeric state T$_{1/2}$=37.2(9) $\mu$s, most likely of spin I=K=6$^{+}$ $\hbar$ \cite{Kallu}. Part of the decays of the short-lived state are missed (Fig. \ref{fig.decaytime}) because the half-life is comparable with the average flight time t=2.19(15) $\mu$s through the separator (from the target to the implantation DSSD). The fraction of produced nuclei, which decay in flight, is estimated to be (28(2))\%. Moreover, a large difference in the relative populations of the ground and isomeric states is measured at low and high beam energies (Fig. \ref{fig.decaytime}). The Fig. \ref{fig.ratio} shows the variation of the ratio of the isomeric state to the total population with beam energy. As pointed out by Heßberger et al. \cite{Hessberger}, in the case of the relative populations of the long-lived and short-lived isomers in $^{254}$No, this behavior is attributed to the contribution of higher partial waves to the fusion cross-section. Unfortunately, the statistics collected at the 235 MeV beam energy did not allow us to extract a reliable ratio at that energy.  Therefore, it is not clear whether the increasing trend of Fig. \ref{fig.ratio} continues or whether a saturation is reached, which would signify that the contribution of even higher partial waves is cut off by fission.

The isotope $^{251}$No was produced in the 1n channel of the reaction $^{48}$Ca with $^{204}$Pb. An excitation function was measured taking into account the contribution from the 3n channel of the reaction $^{48}$Ca with $^{206}$Pb impurities in the target material. The contributions from the other admixtures ($^{207}$Pb, $^{208}$Pb) in the target were ignored due to their negligible values and low evaporation probability \cite{Belozerov}. Thus, the excitation function deduced from the measured number of ER-$\alpha_{1}$ decays is shown in Fig. \ref{fig.cross} by the solid green triangles. We have observed two $\alpha$-lines E$_{\alpha}$ ($^{251}$No)=8616(13) keV and E$_{\alpha}$ ($^{251m}$No)=8669(11) keV, which are in a good agreement with the results from Ref. \cite{Hessberg}. The cross-section of 1.1(2) nb was measured at 213.4 MeV for the reaction $^{204}$Pb($^{48}$Ca,1n) $^{251}$No.  For the last two points of beam energy, it was possible to indicate only cross-section limit values. \\
The experimentally measured excitation functions for the 1n and 2n channels have quite extended tails. A similar behavior was already observed in our early work \cite{Belozerov} and can be explained in part by the large energy losses and straggling in the rather thick titanium backing.

\section{Summary and conclusion}
Neutron-deficient nobelium isotopes produced in the fusion-evaporation reaction $^{48}$Ca+$^{204}$Pb were investigated at the focal plane of the SHELS separator. The study was performed at different beam energies ranging from 213 to 235 MeV in order to measure the excitation functions of the xn evaporation reaction channels. A new alpha activity (T$_{1/2}$=38.3(2.8) ms, E$_{\alpha}$=9129(22) keV) was observed and unambiguously assigned to the decay of $^{249}$No, produced in the 3n evaporation channel. Given the properties of the decay of $^{253}$Rf to $^{249}$No and $^{249}$No to $^{245}$Fm, the ground state of $^{249}$No is assigned the 5/2$^+$[622] neutron configuration and a tentative decay scheme has been established. The upper limit of the spontaneous fission branch extracted for $^{249}$No (b$_{SF}$=($2.3_{-2.3}^{+4.6})\times$10$^{-3}$) indicates a strong hindrance towards fission and is understood as being due to a large specialisation of the 5/2$^+$[622] orbital. In addition, the EC/$\beta^{+}$ decay mode of $^{245}$Fm was clearly evidenced, confirming previous estimates \cite{Khuyagbaatar}. In the case of $^{250}$No (2n evaporation channel) a strong relative enhancement of the isomeric state population over the ground-state population was observed as a function of incident beam energy.  More data at higher beam energies is required to determine the maximum spin that the compound nucleus can withstand. 
\section{Acknowledgements}
Two of us (B. Andel and S. Antalic) are supported by Slovak Research and Development Agency (Contract No. APVV-18-0268) and Slovak Grant Agency VEGA (Project 1/0651/21).

\label{sec:2}

%

%
%

%
%

\end{document}